\documentclass{article}

\usepackage{arxiv}

\usepackage[utf8]{inputenc} % allow utf-8 input
\usepackage[T1]{fontenc}    % use 8-bit T1 fonts
\usepackage{hyperref}       % hyperlinks
\usepackage{url}            % simple URL typesetting
\usepackage{booktabs}       % professional-quality tables
\usepackage{amsfonts}       % blackboard math symbols
\usepackage{nicefrac}       % compact symbols for 1/2, etc.
\usepackage{microtype}      % microtypography
\usepackage{lipsum}
\usepackage{graphicx}
\graphicspath{ {./} }
\usepackage{amsmath}
\usepackage{braket}
\usepackage{hyperref}
\usepackage{amssymb}
\usepackage{subcaption}
\usepackage{amsthm}
\usepackage{mdframed}
\usepackage{tikz}
\usetikzlibrary{positioning, shapes, fit, arrows.meta, calc, decorations.pathreplacing}
\usepackage{mdframed}
\usepackage{comment}
\usepackage{xeCJK}
\usepackage{enumitem}

\newmdtheoremenv{proposition}{Proposition}

\title{Collision models: Markovian and Non-Markovian impurity models}

\author{
Jacob Werner \\
  Department of Physics, The University of Tokyo\\
  5-1-5 Kashiwanoha, Kashiwa-shi, Chiba 277-8574, Japan\\
  \texttt{jacob.werner29@gmail.com}
}

\begin{document}
\maketitle
\begin{abstract}

Collision models provide a flexible framework for studying open quantum systems and introducing non-Markovianity. We present a novel class of collision models, which we term Markovian impurity models, motivated by physical scenarios in which information backflow is periodically blocked. Using the Breuer--Laine--Piilo (BLP) measure of non-Markovianity, we show that if a $p$-impurity model is non-Markovian, then the corresponding $(p+1)$-impurity model is also non-Markovian, implying the existence of a critical $p$ at which the process transitions from Markovian to non-Markovian. Numerical simulations are used to investigate how this critical parameter depends on the system--ancilla and ancilla--ancilla interaction strengths. We find that, altough exceptions exist, increasing either interaction strength generally lowers the critical $p$, while sufficiently weak ancilla--ancilla interactions lead to a regime in which the dynamics remain Markovian even in the absence of impurities. Finally, we briefly introduce the complementary non-Markovian impurity model.

\end{abstract}

% keywords can be removed
\section{Introduction}

Collision models (CMs) are widespread in open quantum systems research, because despite their simplicity, they allow for the characterization of a broad range of phenomena. In particular, they provide a flexible framework for introducing non-Markovianity through many distinct mechanisms \cite{Ciccarello_2022}. Many realistic environments possess memory, and understanding the effects of this is important both fundamentally and for future quantum technologies. Developing simple models in which the strength and origin of non-Markovianity can be systematically varied is thus useful for understanding when and how environmental memory becomes important.

In a CM, rather than dealing with a complicated environment with many degrees of freedom, the environment is modelled as a series of ancillas that sequentially interact with the system. These interactions are called collisions, and each collision designates a discrete time step. The primary contribution of this paper is the introduction and analysis of the Markovian impurity model, a novel class of CMs that may arise naturally in certain physical settings where information backflow is periodically blocked. For these models, the degree of non-Markovianity depends on an impurity parameter $p$. The aim of this paper is to introduce and explain various effects of varying this parameter.
%That is, when we say an event occurs at $t=1$, we mean that the event occurs during or directly after the first collision, depending on context. 

The paper is structured as follows. Sec.~\ref{sec:mark} reviews Markovian CMs. Sec.~\ref{sec:nonmark} discusses how ancilla--ancilla ($YY$) interactions allow for information to flow back to the system. Sec.~\ref{sec:impurity} introduces the Markovian impurity model and presents numerical results, including those concerning the notion of a critical impurity frequency at which a process becomes non-Markovian. We also briefly mention the non-Markovian impurity model. We then conclude and summarize the work in Sec.~\ref{sec:conc}.

%note that maybe i should go back and add subscripts to all Us to designate which y acting on
\section{Markovian collision models}
\label{sec:mark}
\begin{figure}
\centering
\includegraphics[scale=0.6]{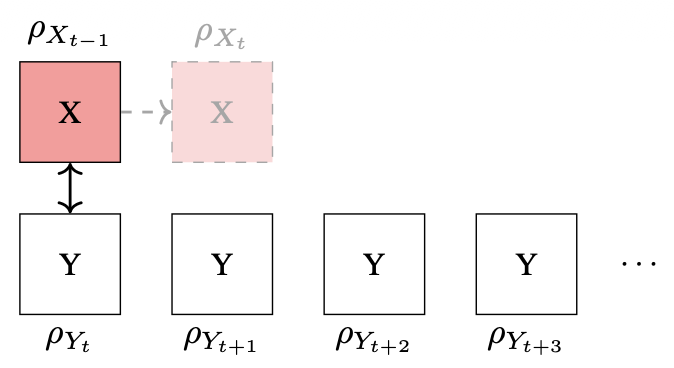}
\caption{Markovian CM schematic}
\label{fig:schematic1}
\end{figure}

In a Markovian CM, each incoming ancilla ($Y$) is initially uncorrelated with both the system ($X$) and other ancillas. A schematic for this process is shown in Fig.~\ref{fig:schematic1}. The state of the system at time $t$, i.e., following the $t$th collision, is given by

\begin{equation}
\rho_{X_t} = \mathcal{E}_t[\rho_{X_{t-1}}] = \operatorname{Tr}_{Y_t}\big[ U_{X_{t-1}Y_t}  (\rho_{X_{t-1}} \otimes \rho_{Y_t} ) U_{X_{t-1}Y_t}^\dagger \big],
\label{eq:markcm}
\end{equation}

\noindent where $\rho_{Y_t}$ is the $t$th ancilla, and $U_{X_{t-1}Y_t}$ is the unitary governing the $t$th system-ancilla ($XY$) interaction. Therefore, Markovian CMs are just repeated completely positive and trace preserving (CPTP) maps, which can be seen by taking the spectral decomposition of the ancilla \cite{Rivas_2012}.

\begin{comment}
\begin{center}
\begin{tikzpicture}[
    every node/.style={font=\small},
    every label/.style={font=\normalsize},
    system/.style={
        draw,
        fill=red!40,
        minimum size=1cm
    },
    ancilla/.style={
        draw,
        minimum size=1cm
    },
    future system/.style={
        draw=black!35,
        dashed,
        fill=red!15,
        text=black!35,
        minimum size=1cm
    }
]

  % System and ancillas
  \node[system, label=above:\(\rho_{X_{t-1}}\)] (S) {X};

  \node[ancilla, below=0.5cm of S,
        label=below:\(\rho_{Y_t}\)] (A1) {Y};

  \node[ancilla, right=0.5cm of A1,
        label=below:\(\rho_{Y_{t+1}}\)] (A2) {Y};

  \node[ancilla, right=0.5cm of A2,
        label=below:\(\rho_{Y_{t+2}}\)] (A3) {Y};

  \node[ancilla, right=0.5cm of A3,
        label=below:\(\rho_{Y_{t+3}}\)] (A4) {Y};

  % Future position of X
  \node[
      future system,
      above=0.5cm of A2,
      label={[text=black!35]above:\(\rho_{X_t}\)}
  ] (Sfuture) {X};

  % Dots
  \node[right=0.3cm of A4] {\(\cdots\)};

  % Interaction 1
  \draw[<->, thick]
      (S) -- (A1);

  % Faded dashed arrow to the future position of X
  \draw[->, dashed, thick, draw=black!35]
      (S.east) -- (Sfuture.west);

\end{tikzpicture}
\end{center}
\end{comment}

We will only consider all-qubit CMs, where the system and each ancilla is a qubit. Also, we will restrict ourselves to time-independent unitaries and incoming ancillas, i.e., $U_{X_{t-1}Y_t}=U_{XY}$, $\rho_{Y_t}=\rho_Y$, and $\mathcal{E}_t=\mathcal{E}$. Additionally, all interactions are described by partial swap operations parameterized by $\theta$, of the form

\begin{equation}
U(\theta) = e^{i \theta S}=\cos(\theta)\,\mathbb{I} + i \sin(\theta)\,S,
\end{equation}

\noindent where $\mathbb{I}$ denotes the $4 \times 4$ identity operator, and $S$ is the SWAP operator. In the computational basis ($\sigma_z$), this is given by

\begin{equation}
S =
\begin{pmatrix}
1 & 0 & 0 & 0 \\
0 & 0 & 1 & 0 \\
0 & 1 & 0 & 0 \\
0 & 0 & 0 & 1
\end{pmatrix}.
\end{equation}

\noindent
The partial swap on a couple states is shown below:
\begin{align}
U(\theta) \, \lvert 00 \rangle &= e^{i\theta} \lvert 00 \rangle, \\
U(\theta) \, \lvert 10 \rangle &= \cos(\theta)\, \lvert 10 \rangle + i \sin(\theta)\, \lvert 01 \rangle.
\end{align}

%maybe rephrase that firstly explanation, but it's not bad
\noindent When $\theta=0$, there is no interaction, and when $\theta=\pi/2$, the interaction is a full swap. 

\section{Non-Markovian collision models}
\label{sec:nonmark}
\begin{comment}
While the definition of classical Markovianity is well-defined through conditional probabilities, there is no clear consensus on the definition for quantum processes. However, all definitions aim to capture some concept of memory effects or information backflow from the environment. It is often said that in a Markovian process,
the system continually loses information to the environment, while there is an exchange of
information in a non-Markovian process \cite{Breuer_2016}. The information that goes to the environment
and later comes back is stored in correlations or changes to the environmental states.
\end{comment}

In this section, we introduce the basics of how ancilla--ancilla ($YY$) interactions give rise to non-Markovian CMs, and in Sec.~\ref{sec:impurity}, we will introduce the Markovian impurity model, a specific way of varying the degree of Markovianity.

There is no clear consensus on the defintion of Markovianity for quantum processes, but throughout this paper, we will employ the Breuer-Laine-Piilo (BLP) measure of non-Markovianity, which is commonly used for discrete-time dynamical processes. Introduced in Ref.~\cite{Breuer_2009}, the measure is based on the principle that Markovian dynamics cannot increase the distinguishability between two states. This distinguishability is quantified by the trace distance, which for two states $\rho_1$ and $\rho_2$ is defined as

\begin{equation}
D\left( \rho_1, \rho_2 \right) = \frac{1}{2} \left\| \rho_1 - \rho_2 \right\|_1.
\end{equation}

The BLP measure $\mathcal{N}_{\mathrm{BLP}}$ is obtained by summing all increases in the trace distance:

\begin{equation}
\mathcal{N}_{\mathrm{BLP}}
= \max_{\rho_{1,0}, \rho_{2,0}}
\sum_{t:\, \Delta D > 0}
\Big[
D(\rho_{1,t}, \rho_{2,t})
- D(\rho_{1,t-1}, \rho_{2,t-1})
\Big].
\label{eq:discrete}
\end{equation}

Rather than maximizing over all initial state pairs, we will simply use the $\sigma_x$ eigenstates to determine $\mathcal{N}_{\mathrm{BLP}}$. The difference is negligible for the qualitative conclusions of this paper. Note that for Markovian processes, the trace distance between two states must decrease so that $\mathcal{N}_{\mathrm{BLP}}=0$ \cite{Lindblad1975}.

\subsection{Ancilla--ancilla interactions}
If we allow ancilla--ancilla ($YY$) interactions, there is now a clear means by which information backflow can occur \cite{Ciccarello_2022}. Note that information that goes to the environment and later comes back is stored in correlations or changes to the environmental states. If there is a nearest-neighbor $YY$ interaction after the $XY$ interaction, the state at time $t$, $\rho_{X_t}$, is given by

\begin{align}
\rho_{X_t} &= \mathcal{E}_t\big[\rho_{X_{t-1}}\big]  \nonumber \\
&= \operatorname{Tr}_{Y_t,Y_{t+1}}\big[ U_{YY}U_{XY} \left( \rho_{X_{t-1}Y_{t}} \otimes \rho_{Y_{t+1}}\right) U_{XY}^\dagger U_{YY}^\dagger\big],
\label{eq:nonm}
\end{align}

\noindent where $U_{XY}$ and $U_{YY}$ are parameterized by $\theta_{XY}$ and $\theta_{YY}$, respectively. Before the $XY$ interaction occurs, the system is now in general already correlated to the ancilla. The joint system-ancilla state at time $t$, $\rho_{X_tY_{t+1}}$, is given by the same expression above, but only taking the trace over $Y_t$. This is the joint system-ancilla state where the ancilla is the incoming ancilla. To obtain the joint system-ancilla state where the ancilla is the outgoing ancilla, trace over $Y_{t+1}$. Note that due to the correlations, the map is now, in general, non-CPTP and time-dependent. 

\begin{comment}
Also, the new Hamiltonian is given by

\begin{equation}
H_I(t_0)
= \sum_{t=1}^{\infty} \chi_t(t_0)\, H_{X Y_t}
+ \sum_{t=1}^{\infty} \eta_t(t_0)\, H_{Y_t Y_{t+1}},
\end{equation}

\noindent where $\eta_t(t_0)$ is $1$ during the corresponding $YY$ interaction and $0$ otherwise. Note that if we only consider partial swaps, we still have $[H_I,H_X \;+\; \sum_{t=1}^{\infty} H_{Y_t}]=0$. 
\end{comment}

In this paper, we will only consider thermal qubits given by

\begin{equation}
\rho_{\mathrm{th}}=\frac{e^{-\beta_Y \sigma_z}}{\operatorname{Tr}\!\left[e^{-\beta_Y \sigma_z}\right]},
\end{equation}

%maybe find better citation for thermal operation thing
\noindent where $\beta_Y=1/{k_B T_Y}$ is the inverse temperature of the ancillas. The CM then falls into the restricted class of operations known as thermal operations \cite{_enya_a_2022}. In these cases, $U_{YY}U_{XY}(\rho_{Y}\otimes\rho_Y)U_{XY}^\dagger U_{YY}^\dagger = \rho_{Y}\otimes\rho_Y$, i.e., the global fixed point of the dynamics is $\rho_Y$, and it can be shown that all $\rho_X$ eventually tend to $\rho_Y$ \cite{_enya_a_2022}. Therefore, even in the non-Markovian case where information backflow may occur on small time scales, in the long-time limit, the system will still thermalize for these processes. Since thermal states are diagonal in the $\sigma_z$ basis, for simplicity we will parameterize them by their excited state populations, i.e., $\bra{0}\rho_{\mathrm{th}}(a)\ket{0}=a$ and $\bra{1}\rho_{\mathrm{th}}(a)\ket{1}=1-a$.

\section{Impurity models}
\label{sec:impurity}

\begin{figure}
\centering
\includegraphics[scale=0.6]{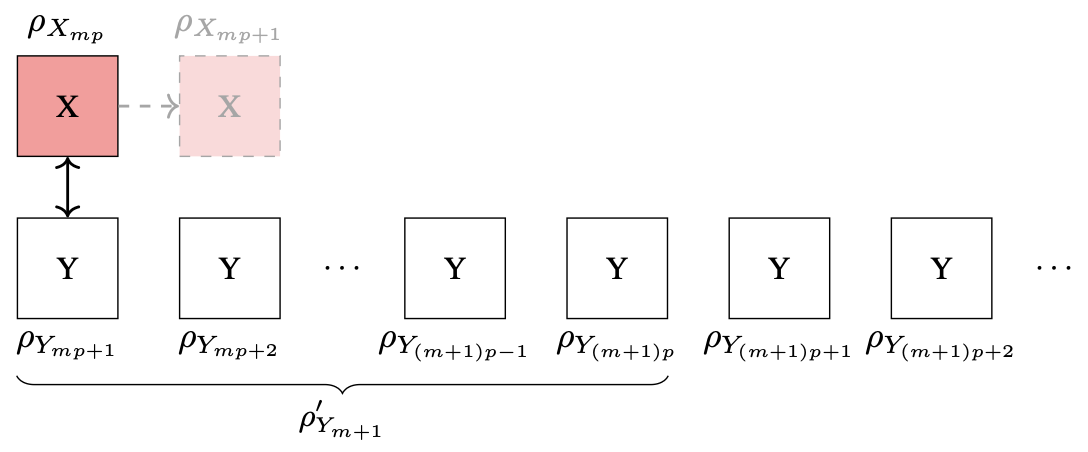}
\caption{Markovian impurity model schematic}
\label{fig:schematic2}
\end{figure}

\subsection{Markovian impurity model}
We first present the Markovian impurity model, which involves inserting ``impurities" at fixed time intervals. Afterwards, we will briefly mention non-Markovian impurities. In the context of Markovian impurity models, we will define a $p$-impurity model as follows. For every $t \in p\mathbb{N}$ the map is given by

\begin{equation}
\rho_{X_t}=\mathcal{E}'_t[\rho_{X_{t-1}}]=\operatorname{Tr}_{Y_t}\big[ U_{XY} \left( \rho_{X_{t-1}Y_{t}}\right) U_{XY}^\dagger\big].
\label{eq:impurity11}
\end{equation}

\noindent For all other time steps, the map is given by

\begin{equation}
\rho_{X_t} 
= \mathcal{E}_t[\rho_{X_{t-1}}]
= \operatorname{Tr}_{Y_t, Y_{t+1}}
\!\big[
U_{Y Y}\, U_{X Y}
\big( \rho_{X_{t-1} Y_{t}} \otimes \rho_{Y_{t+1}} \big)
U_{X Y}^\dagger \, U_{Y Y}^\dagger
\big].
\label{eq:impurity12}
\end{equation}

\noindent We have used $\mathcal{E}'$ and $\mathcal{E}$ to distinguish the two maps. For clarity, see the schematic for a Markovian impurity model in Fig.~\ref{fig:schematic2}. Note that for $t \in p\mathbb{N}+1$, $\rho_{X_{t-1}Y_{t}}$ is an uncorrelated state given by $\rho_{X_{t-1}}\otimes\rho_{Y_t}$, where $\rho_{Y_t}$ is the initial state of the ancilla. Therefore, for $t \in p\mathbb{N}+1$, the map is CPTP. A $p$-impurity model effectively has $U_{YY}=\mathbb{I}$ every $p$ collisions, and hence we would expect that increasing $p$ would increase the degree of non-Markovianity. Also, observe that a $1$-impurity model has an ``impurity" at every time step, i.e., no $YY$ interactions, and hence the model reduces to the Markovian CM of Eq.~\eqref{eq:markcm}.

We aim to show the following:

%make sure calling this a lemma is correct
\begin{proposition}
If a $p$-impurity model is non-Markovian ($\mathcal{N}_{\mathrm{BLP}}>0$), then the corresponding $(p+1)$-impurity model is also non-Markovian. 
\label{lemma}
\end{proposition}

\noindent This can be equivalently stated another way: if a $p$-impurity model is Markovian, then the corresponding $(p-1)$-impurity model is Markovian. Before proving this, we will first make a brief observation on how the coarse-graining of time affects whether or not a process is considered Markovian.
%put a schematic of model right here, that will definitely be useful

\subsubsection{Coarse-graining and non-Markovianity}
The discrete-time BLP measure only considers the system before and after applying a map $\mathcal{E}$, i.e., we wait until $U_{XY}$ (and possibly $U_{YY}$) has been fully applied before considering the system. However, this operation has a duration $\Delta t$. Let us choose units such that this duration is $1$, and assume that the map from $t$ to $t+1$ is CPTP. For $\epsilon \in (0,1)$, the map from $t$ to $t+\epsilon$ is CPTP, but in general, the map from $t+\epsilon$ to $t+1$ is not. Therefore, CP-divisibility is broken, and the trace distance of two states from $t+\epsilon$ to $t+1$ may increase. We now see that whether or not a dynamics is considered Markovian can be determined by how finely we choose to coarse-grain time.

We can consider a similar result for a Markovian impurity model. If we coarse-grain time and only consider the system for $t \in p\mathbb{N}$, then for $m \in\mathbb{N}$, we can write the map from $\rho_{X_{mp}}$ to $\rho_{X_{(m+1)p}}$ as a CPTP map $\mathcal{E}''$ given by

\begin{align}
\rho_{X_{(m+1)p}}
&=\mathcal{E}''\!\big[\rho_{X_{m p}}\big]=\mathcal{E}'_{(m+1)p} \mathcal{E}_{(m+1)p - 1} \cdots \mathcal{E}_{mp+1}\big[\rho_{X_{mp}}\big] \\
&= \operatorname{Tr}_{Y'_{m+1}}
\!\big[
U'_{X Y}
\big( \rho_{X_{m p}} \otimes \rho'_{Y_{m+1}} \big)
U_{X Y}'^{\dagger}
\big],
\end{align}

\noindent where $\rho'_{Y_{m+1}}$ and $U_{X Y_{(m+1)p}}$ are given by

\begin{align}
\rho'_{Y_{m+1}}
&= \bigotimes_{t = m p + 1}^{(m+1)p} \rho_{Y_t}, \\[0.5em]
U'_{X Y}
&=
U_{X Y_{(m+1)p}}\,
U_{Y_{(m+1)p-1}\,Y_{(m+1)p}}\,
U_{X Y_{(m+1)p-1}}\,
\cdots\,
U_{Y_{m p+1}\,Y_{m p+2}}\,
U_{X Y_{m p+1}}.
\end{align}

\noindent 
The time dependence of the unitaries is included to clarify which ancillas they act upon. Note that the partial trace over an ancilla commutes with all operators that do not act on that ancilla, which is why we can wait to trace out all $p$ ancillas at $t=(m+1)p$. Therefore, we have shown that by coarse-graining time and considering $\bigotimes_{t=mp+1}^{(m+1)p} \rho_{Y_t}$ as one larger ancilla, the $p$-impurity model can be viewed as a Markovian map, i.e., the trace distance between states never increases for $t \in p\mathbb{N}$.

However, in this paper we will concern ourselves with monitored CMs where measurements are made at every time step, and hence the system must be considered for all $t \in \mathbb{N}$. Therefore, for our purposes, the map is not CP-divisible and may satisfy $\mathcal{N}_{\mathrm{BLP}}>0$ for $p\geq2$.

\subsubsection{Critical $p$: Definition}
In order to prove Prop.~\ref{lemma}, we first show that the reduced system state is the same before and after a $YY$ interaction, as shown below:

\begin{align}
\rho_{X_t}
&= \operatorname{Tr}_{Y_t, Y_{t+1}}
\!\big[
U_{Y Y}U_{X Y}
\big( \rho_{X_{t-1} Y_{t}} \otimes \rho_{Y_{t+1}} \big)
U_{X Y}^\dagger \, U_{Y Y}^\dagger
\big] \nonumber\\
&= \operatorname{Tr}_{Y_t, Y_{t+1}}
\!\big[
U_{X Y}
\big( \rho_{X_{t-1} Y_{t}} \otimes \rho_{Y_{t+1}} \big)
U_{X Y}^\dagger \, 
U_{Y Y}^\dagger U_{Y Y}
\big] \nonumber\\
&= \operatorname{Tr}_{Y_t}
\!\big[
U_{X Y}\, \rho_{X_{t-1} Y_{t}} \, U_{X Y}^\dagger
\big].
\label{partial}
\end{align}

\noindent In principle, to determine if $\mathcal{N}_{\mathrm{BLP}}>0$, we must check whether there exists any pair of initial states $\{\rho_{1,0}, \rho_{2,0}\}$ and any $t$ where $\Delta D(t)>0$, i.e., where 

\begin{equation}
\Delta D(t)=D(\rho_{1,t}, \rho_{2,t}) - D(\rho_{1,t-1}, \rho_{2,t-1}) > 0.
\end{equation}

%$t \in p\mathbb{N}+1$ i say at this time uncorrelated to the incoming ancilla, that's correct, but may be clearer to say at this time $t \in p\mathbb{N}$ Note that I am saying time step 3 is from state 2 to 3

Say we find a pair $\{ \rho_{1,0}, \rho_{2,0}\}$ such that $\Delta D(T)>0$, and $T=mp+k$, where $m,k \in \mathbb{N}$ and $1< k\leq p$. At the $t$th time step, where $t \in p\mathbb{N}+1$, $\rho_X$ is uncorrelated to the incoming ancilla ($\rho_{X_{t-1}Y_{t}}=\rho_{X_{t-1}}\otimes\rho_{Y_t}$). Therefore, we can take $\{\rho'_{1,0}, \rho'_{2,0}\}=\{\rho_{1,mp}, \rho_{2,mp}\}$, and we will have $\Delta D(k)>0$. Therefore, we only need to check a maximum of $p$ time steps to determine if $\mathcal{N}_{\mathrm{BLP}}>0$, assuming that we check all possible initial state pairs (or use suitable heuristics).

Now Prop.~\ref{lemma} becomes almost trivial. If a $p$-impurity model is determined to be non-Markovian, then we have found an initial pair $\{\rho_{1,0}, \rho_{2,0}\}$ such that $\Delta D(k)>0$ for $1< k\leq p$. Consider passing the same pair through the $(p+1)$-impurity model (call the pair $\{\rho'_{1,0}, \rho'_{2,0}\}$). For $t\leq p-1$, we clearly have $\{\rho'_{1,t}, \rho'_{2,t}\}=\{\rho_{1,t}, \rho_{2,t}\}$. Also, on the $p$th time step, $U_{Y_p Y_{p+1}}$ is a partial swap for the $(p+1)$-impurity model and effectively $\mathbb{I}$ for the $p$-impurity model, and Eq.~\eqref{partial} shows that $\{\rho'_{1,p}, \rho'_{2,p}\}=\{\rho_{1,p}, \rho_{2,p}\}$ still holds. Hence, for any valid $k$, the $(p+1)$-impurity model satisfies $\Delta D(k)>0$ and is therefore also non-Markovian. 

A process is trivially Markovian for $p=1$. Therefore, if a process is non-Markovian for $p=\infty$, then at some finite value there was a transition from Markovian to non-Markovian. We call this value, the smallest value for which the $p$-impurity model is non-Markovian, the critical $p$, denoted as $p'$.

%sort of intuitive that increasing p slowly increases correlations with environment, which makes map less CP-divisible, hence more non-Markovian. here talking about correlations before, but should be related to correlations after

%It was mentioned that there is a tradeoff related to increasing $\theta_{XY}$; high values should cause earlier terms in Eq.~\eqref{eq:discrete} to be larger, but there will generally be fewer terms due to the system reaching the steady state more quickly. However, here we are not interested in the magnitude of $\mathcal{N}_{\mathrm{BLP}}$, but rather only whether it is nonzero. Therefore, only the first effect is relevant here, as we only require one nonzero term. 

\subsubsection{Critical $p$: Numerical results}
%graph p-1 not p-2, so we can see the horizontal line, maybe just graph p for simplicity
%do heatmap of critical p as function of thetaSA and thetaAA
%also graph asymptote thetaAA as function of thetaSA
%don't call it a prediction, call it a sketch for clarity
%do what hatano said, since increasing p is discrete, increase thetaAA

%maybe mention the thing with theta_XY, may not be worth it to mention it though.
In this section, we numerically investigate $p'$ for the canonical non-Markovian CM described by Eq.~\eqref{eq:nonm}, with $XY$ and $YY$ interactions parameterized by $\theta_{XY}$ and $\theta_{YY}$, respectively. Starting with the $1$-impurity model with no $YY$ interactions and increasing $p$, the simulations will determine if and when the model transitions to being non-Markovian, where $\mathcal{N}_{\mathrm{BLP}}$ is calculated using the $\sigma_x$ eigenstates as the initial state pair. We expect that increasing $\theta_{XY}$ and $\theta_{YY}$ will tend to increase $\mathcal{N}_{\mathrm{BLP}}$ and decrease $p'$. 

\begin{figure}
\centering
\includegraphics[scale=0.6]{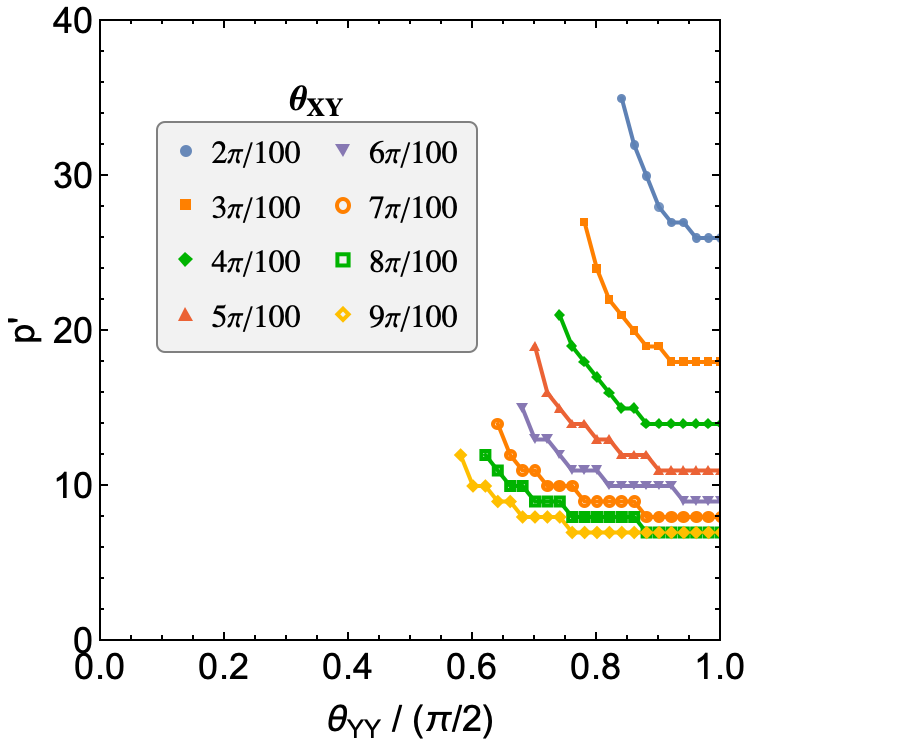}
\caption{$p'$ is plotted against $\theta_{YY}$ for different values of $\theta_{XY}$, with $\rho_Y=\rho_{\mathrm{th}}(0.3)$. The BLP measure $\mathcal{N}_{\mathrm{BLP}}$ was calculated using the $\sigma_x$ eigenstates as the initial state pair. The curves approach an asymptote as $\theta_{YY}$ is decreased.}
\label{fig:critp}
\end{figure}

\begin{figure}
\centering
\includegraphics[scale=0.6]{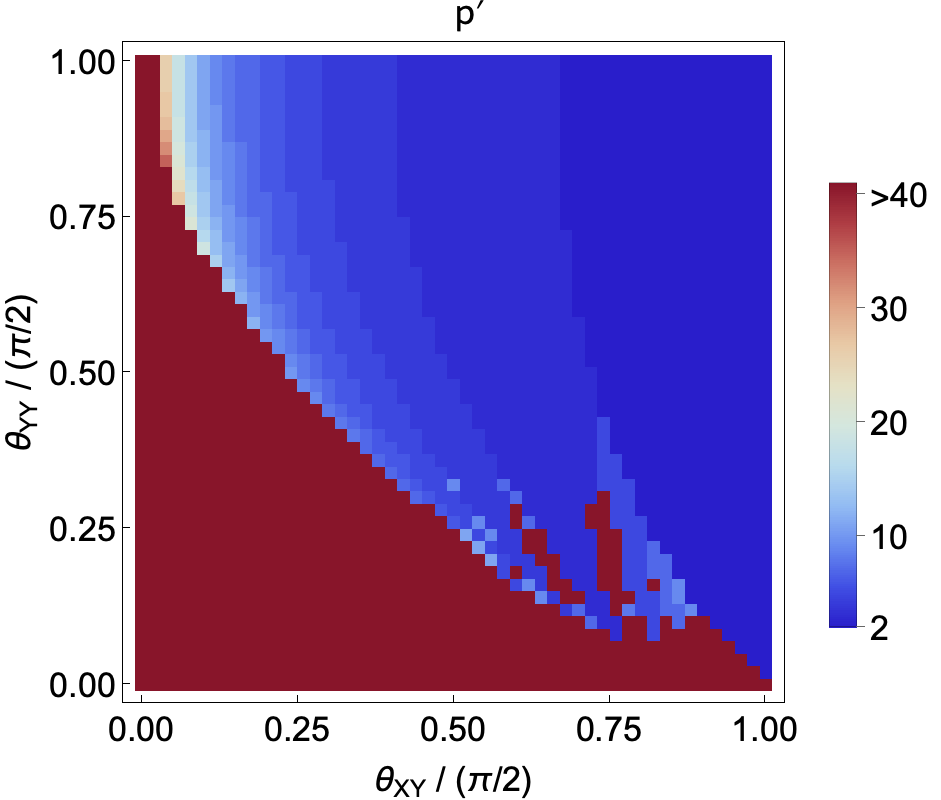}
\caption{$p'$ is plotted as a function of $\theta_{XY}$ and $\theta_{YY}$, with $\rho_Y=\rho_{\mathrm{th}}(0.3)$. Due to computational constraints, we only checked up to $p=40$ for non-Markovianity. Therefore, $p'>40$ could correspond to a genuine $p'>40$ or that the process will remain Markovian even for $p=\infty$.}
\label{fig:heatplot}
\end{figure}

%if in the simulations still exist that fixed Xy, decreasing YY decreases p, then include that, may have just been oddity though
%give detailed information on behavior in certain ranges once you get the data, refer to other latex for guide. discuss other weird numerical results once you have them
Numerical results of $p'$ as a function of $\theta_{YY}$ for various values of $\theta_{XY}$ are shown in Fig.~\ref{fig:critp}. Remember that for a specific $\theta_{XY}$ curve, on or above the line is non-Markovian ($\mathcal{N}_{\mathrm{BLP}}>0$). We can see that for small $\theta_{XY}$, $p'$ is large, even for $\theta_{YY}\approx\pi/2$. As expected, in general for a fixed $\theta_{XY}$, decreasing $\theta_{YY}$ results in a greater $p'$, and similarly, for a fixed $\theta_{YY}$, decreasing $\theta_{XY}$ usually results in a greater $p'$. We also show $p'(\theta_{XY},\theta_{YY})$ in Fig.~\ref{fig:heatplot} for clarity, and it is apparent that the relationship is nontrivial. 

Recall that due to the discrete nature of $p$, these curves are all step functions. Also observe that, for high enough $\theta_{XY}$ and $\theta_{YY}$, even $p=2$ results in $\mathcal{N}_{\mathrm{BLP}}>0$. Additionally, it is apparent that for every value of $\theta_{XY}$, an asymptote is eventually approached when decreasing $\theta_{YY}$, and to the left of the asymptote, the process remains Markovian for all $p$ values, even when $p=\infty$, i.e., no impurities. 

Note that only values of $p'$ up to 40 were tested due to computational constraints. However, within 40 time steps, the system is in most cases very close to equilibrium, and revivals in trace distance are therefore unlikely after that. Also, when we examined values of $\theta_{YY}$ close to the asymptote by looking at 100 time steps, we observed a very quick jump from values of $p'$ being on the order of 10 to being greater than 100 (likely infinite). Additionally, although Ref.~\cite{McElvogue_2026} only considered a purely non-Markovian CM (effectively $p=\infty$), they observed a similar rapid asymptotic approach to $\mathcal{N}_{\mathrm{BLP}}=0$ upon decreasing $\theta_{YY}$. Hence, we can confidently say that the $p'$ values of Fig.~\ref{fig:heatplot} which are greater than 40 most likely correspond to the process remaining Markovian even for $p=\infty$.

\begin{figure}
\centering
\includegraphics[scale=0.6]{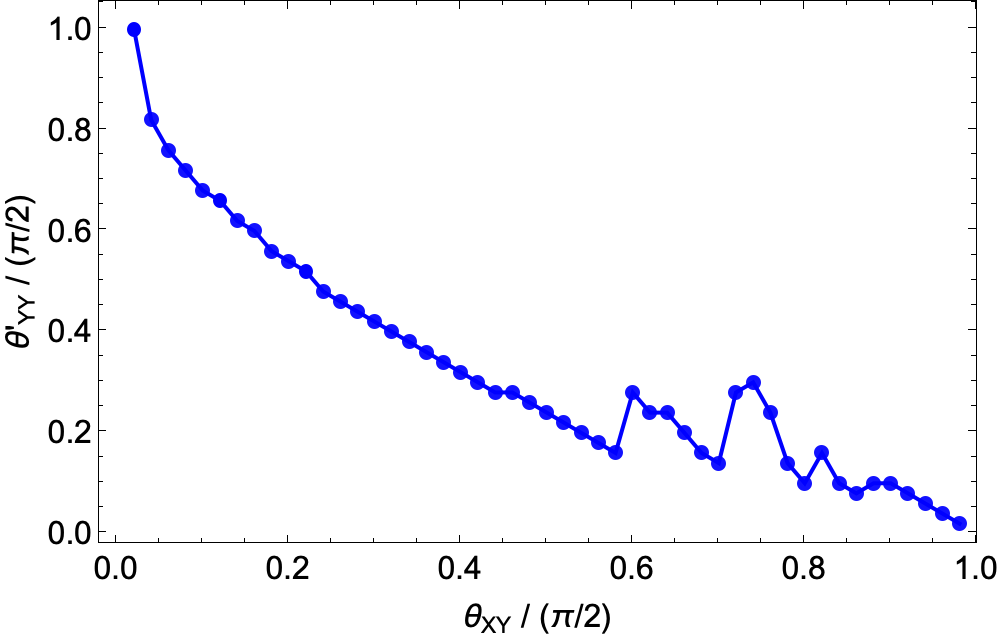}
\caption{The maximum value of $\theta_{YY}$ for which $p=\infty$ still satisfies $\mathcal{N}_{\mathrm{BLP}}=0$ is plotted against $\theta_{XY}$.}
\label{fig:asymptote}
\end{figure}

The functional relationship of the value of $\theta_{YY}$ at which the asymptote occurs, $\theta'_{YY}(\theta_{XY})$, is shown in Fig.~\ref{fig:asymptote}, and we can see that as $\theta_{XY}$ is increased, $\theta'_{YY}$ generally decreases. This  results in a downward and to the left shift of the $p'$ curves of Fig.~\ref{fig:critp} as $\theta_{XY}$ is increased.

%change this later, cuz maybe we don't, go back after discussing the symmetric plots
Observe that at higher values of $\theta_{XY}$, there is a complex interplay between correlations, changes to environmental states, and the rate of approach to equilibrium. Strange phenomena can occur, such as increasing $p'$ upon increasing $\theta_{YY}$ (see Fig.~\ref{fig:heatplot}) or increasing $\theta'_{YY}$ upon increasing $\theta_{XY}$ (see Fig.~\ref{fig:asymptote}). 

\begin{comment}
%maybe fix to make it p, not p-2
\begin{figure}[ht]
\centering
\begin{tikzpicture}[scale=0.6]
  % Axes
  \draw[->] (-0.5, 0) -- (6.5, 0) node[right] {$\theta_{YY}$};
  \draw[->] (0, -0.5) -- (0, 10) node[above] {$p'-2$};

  % Labels
  \node at (6, 9) {\footnotesize $\theta_{XY}\approx0$};
  \node at (1.9, 2) {\footnotesize $\theta_{XY}\approx\frac{\pi}{2}$};

  % Curves
  \draw[thick, blue] (6,7) .. controls (5.25,7) and (5.25,8.5) .. (5.25,9);
  \draw[thick, red] (6,5.5) .. controls (5.0,5.5) and (5.0,8.5) .. (5.0,9);
  \draw[thick, green!70!black] (6,4) .. controls (4.75,4) and (4.75,8.5) .. (4.75,9);
  \draw[thick, orange] (6,2.5) .. controls (4.5,2.5) and (4.5,8.5) .. (4.5,9);
  \draw[thick, purple] (6,1) .. controls (4.25,1) and (4.25,8.5) .. (4.25,9);
  \draw[thick, cyan] (6,0) .. controls (4.0,0) and (4.0,8.5) .. (4.0,9);
  \draw[thick, magenta] (5,0) .. controls (3.2,0) and (3.2,8.5) .. (3.2,9);
  \draw[thick, teal] (3.5,0) .. controls (2.0,0) and (2.0,8.5) .. (2.0,9);
  \draw[thick, brown] (2,0) .. controls (0.9,0) and (0.9,8.5) .. (0.9,9);
\end{tikzpicture}

\caption{Schematic of $p'$ as a function of $\theta_{YY}$ for various $\theta_{XY}$ values, to aid understanding}
\label{fig:pred}
\end{figure}

The actual curves are step functions, and numerical noise and multiple plots make the relationships a bit hard to graphically read. Therefore, to aid clarity for the reader and emphasize the qualitative relationships, a rough schematic with continuous curves is shown in \ref{fig:pred}.
\end{comment}

\subsection{Non-Markovian impurity model}
We briefly introduce the concept of a non-Markovian $p$-impurity model, which we define as follows. For $t \in p\mathbb{N}$ the map is given by

\begin{equation}
\rho_{X_t} 
= \mathcal{E}_t[\rho_{X_{t-1}}]
= \operatorname{Tr}_{Y_t, Y_{t+1}}
\!\big[
U_{Y Y}\, U_{X Y}
\big( \rho_{X_{t-1} Y_{t}} \otimes \rho_{Y_{t+1}} \big)
U_{X Y}^\dagger \, U_{Y Y}^\dagger
\big].
\end{equation}

\noindent For all other time steps, the map is given by

\begin{equation}
\rho_{X_t}=\mathcal{E}'_t[\rho_{X_{t-1}}]=\operatorname{Tr}_{Y_t}\big[ U_{XY}  \rho_{X_{t-1}Y_{t}} U_{XY}^\dagger\big].
\end{equation}

\noindent By construction, $\rho_{X_{t-1} Y_{t}}$ can only be a correlated state for $t \in p\mathbb{Z}^++1$. The non-Markovian impurity model is essentially the opposite of the Markovian impurity model given by Eqs.~\eqref{eq:impurity11}--\eqref{eq:impurity12}, i.e., the map is a CPTP map except for $t \in p\mathbb{Z}^++1$. In order to have $\mathcal{N}_{\mathrm{BLP}}>0$, we require that just one $YY$ interaction can build up enough correlations for the trace distance between two states to increase.

%could mention the reduced Hilbert space thing too, not sure if necessary though, also could mention how critical p concept not as relevant, only for reduced hilbert space thing. if reduced hilbert space doesn't really matter, most models cross over at p=inf, 1, or never, to non-Markovian.

Consider, for example, the Markovian $10$-impurity model. The model has $9$ collisions involving $YY$ interactions to build up correlations before a Markovian impurity, but for the non-Markovian impurity model, unless we are in a 1-impurity model, we only have one collision involving a $YY$ interaction to build up correlations before the next CPTP map. Therefore, we expect this model to have $\mathcal{N}_{\mathrm{BLP}}>0$ only for very high $\theta_{YY}$ and $\theta_{XY}$. For this reason, this model is considerably less interesting and only mentioned here for completeness.

\section{Conclusion}
\label{sec:conc}
The overarching goal of this paper is to introduce a novel class of non-Markovian CMs that may arise in certain physical scenarios. For example, if ancillas approach the system in groups of $p$ ancillas, information backflow may occur within the group but not between groups. The effect on the degree of non-Markovianity is non-trivial and has been explored throughout this paper. 

We showed that if a $p$-impurity model is non-Markovian by the BLP measure, then the corresponding $(p+1)$-impurity model is also non-Markovian. This suggests a crossover $p$ at which the process becomes non-Markovian, and we call this the critical $p$. We presented numerical results that show that in general increasing $\theta_{XY}$ or $\theta_{YY}$ reduces, or leaves unchanged, the critical $p$. We also showed that upon decreasing $\theta_{YY}$, an asymptote is approached beyond which the process remains Markovian even for $p=\infty$. We then briefly mentioned the non-Markovian impurity model. Although several interesting features of Markovian impurity models have been identified, a systematic exploration of the model lies beyond this paper's scope and remains an open direction for future work.

\bibliographystyle{ieeetr}
\bibliography{citations}

\end{document}